# Electronic Structures of Graphene Layers on a Metal Foil: The Effect of Point Defects

Hui Yan[1,§], Cheng-Cheng Liu[2,3,§], Ke-Ke Bai[1,§], Xuejiao Wang[2], Mengxi Liu[4], Wei Yan[1], Lan Meng[1], Yanfeng Zhang[4,5], Zhongfan Liu[4], Rui-fen Dou[1], Jia-Cai Nie[1], Yugui Yao[2,a], and Lin He[1,b,*]

[1] Department of Physics, Beijing Normal University, Beijing, 100875, People's Republic of China
[2] School of Physcis, Beijing Institute of Technology, Beijing 100081, People's Republic of China
[3] Beijing National Laboratory for Condensed Matter Physics and Institute of Physics, Chinese Academy of Sciences, Beijing 100190, People's Republic of China
[4] Center for Nanochemistry (CNC), College of Chemistry and Molecular Engineering, Peking University, Beijing 100871, People's Republic of China
[5] Department of Materials Science and Engineering, College of Engineering, Peking University, Beijing 100871, People's Republic of China

Here we report a facile method to generate a high density of point defects in graphene on metal foil and show how the point defects affect the electronic structures of graphene layers. Our scanning tunneling microscope measurements, complemented by first-principles calculations, reveal that the point defects result in both the intervalley and intravalley scattering of graphene. The Fermi velocity is reduced in the vicinity area of the defect due to the enhanced scattering. Additionally, our analysis further points out that periodic point defects can tailor the electronic properties of graphene by introducing a significant bandgap, which opens an avenue towards all-graphene electronics.

The fact that quasiparticles in graphene mimic massless Dirac fermions is a consequence of graphene's bipartite honeycomb lattice, which consists of two equivalent carbon sublattices (they are viewed as sublattice pseudospin) [1-9]. The graphene's unique crystal structure results in linear energy dispersion near the Fermi energy and two independent Dirac cones centered at the opposite corners of the Brillouin zone, commonly called K and K′. The two Dirac cones, which are mathematically similar to electron spin, are treated as valley isospin and suggested as carriers of information [10-13]. Nevertheless, graphene is not immune to defect. Any defect that deforms the structure of the original honeycomb lattice has a strong impact in the electronic properties of graphene [14-22]. For example, lattice deformation of graphene will introduce effective gauge fields and influence the Dirac fermions in graphene like an effective magnetic field [14,15,23]. Point defect, which is on the order of the lattice spacing of graphene, could provide a large momentum transfer of the Dirac fermions and lead to scattering from K to K′, or vice versa [17,20].

In this Letter, we report a facile method to introduce a high density of point defects in graphene on metal foil by thermal annealing. Effects of the point defects on the electronic structures of graphene layers are studied carefully by scanning tunneling microscopy and spectroscopy (STM and STS). The point defects result in both the intervalley and intravalley scattering of graphene and lead to a reduction of the Fermi velocity in the vicinity area of the defect. Our result indicates that it's possible to realize all-graphene electronics as soon as that the point defects can be patterned into graphene in a controllable way.

There are various atomic-scale defects, such as heptagon-pentagon topological defects, adatoms, dopants, atomic vacancies, in graphene layers [17,20-22,24-28]. These point defects can spontaneous appear at the stage of the graphene growth [20] and can be deliberately introduced by irradiation [22,29], resonance plasma [27], and chemical treatment [21]. Here we report a new and facile method to introduce a high density of point defects in graphene. In our experiment, the graphene sample was grown on a polycrystalline Rh foil via a traditional ambient pressure chemical vapor deposition (CVD) method. The sample was synthesized at 1000 ºC [16,30,31]. The graphene growth on the Rh foil was attributed to a segregation mechanism [31]. Briefly, methane gases decomposed on the Rh foil at 1000 ºC and the carbon atoms dissolve into the Rh foil because of its high carbon solubility at high temperature. In the cooling process, the carbon atoms would segregate from the Rh foil to the surface, forming graphene layers.

We will show subsequently that the temperature dependent carbon solubility of Rh foils can be used to introduce point defects in graphene. Unlike graphene on single-crystal Rh [32-34], the coupling between graphene and the Rh foil is very weak. The as-grown graphene sample is almost point defect free within each domain (a few microns in size) [16,30,31]. The tunneling spectrum of the as-grown graphene monolayer is almost identical to that of freestanding graphene monolayer (see supplementary material [35] for details of STM measurement and Fig. S1 of for STM images and STS of the as-grown graphene monolayer). To generate point defects, the temperature of the as-grown sample was increased from room-temperature to about 300 ºC and the sample was kept at this temperature for several hours in ultrahigh vacuum condition. The carbon solubility of Rh foil increases with temperature and a small partial of carbon atoms in graphene re-dissolve into Rh foil at positions where the graphene and the Rh foil are point-contact (see Fig. S2 [35]). When the sample was cooled

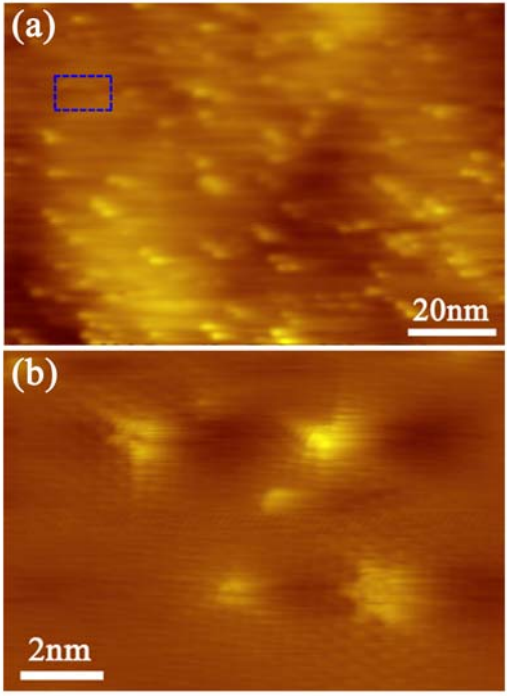

FIG. 1 (color online). (a) A STM image of graphene monolayer with high density of point defects ($V_{sample}$ = 580 mV and I = 10.3 pA). The bright positions are the point defects. (b) Zoom-in image of several typical point defects, including single-atom vacancies and a flower defect, in the blue frame of panel (a).

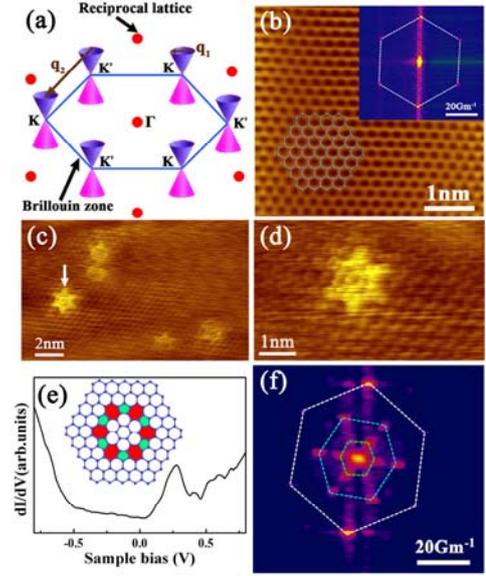

FIG. 2 (color online). (a) The reciprocal lattice, first Brillouin zone, and schematic Dirac cones of graphene monolayer. The wave vectors $q_1$ and $q_2$ represent intravalley and intervalley scattering respectively. (b) Atomic-resolution image of defect free graphene monolayer ($V_{sample}$ = 600 mV and I = 14.8 pA). The atomic structure of graphene is overlaid onto the STM image. The inset is fast Fourier transform of the main panel. (c) A STM image of graphene monolayer with several flower defects ($V_{sample}$ = 416 mV and I = 8.01 pA). (d) Zoom-in image of the flower defect as pointed out by the arrow in panel (a). A clear $\sqrt{3}\times\sqrt{3}R30°$ interference pattern is observed in proximity of the defect. (e) A typical tunneling spectrum recorded on the flower defect. The inset shows schematic structure of a flower defect. (f) Fast Fourier transform of a STM image around a flower defect. The outer six spots correspond to the reciprocal lattice of graphene. The middle six spots arise from the intervalley scattering. The inner six spots are attributed to the reciprocal lattice of the flower defect. The center bright region of the fast Fourier transform image is related to the intravalley scattering.

down again, the re-dissolved carbon atoms re-segregate from bulk to surface. The reconstructed structure of the sample could possibly be the original honeycomb lattices, then, we will still obtain defect free sample. However, the reconstructed structure is more likely to introduce point defects, including topological defects, atomic vacancies, and adatoms, in the graphene sample. As a consequence, we obtained graphene with a high density of point defects after the thermal annealing, as shown in Fig. 1(a).

The type of the point defects depends on a balance between the re-dissolution and re-segregation of carbon atoms in a local position. For the case $N_d = N_s$ (here $N_d$ and $N_s$ are the number of re-dissolved and re-segregated carbon atoms in a local position respectively), the generated point defect by the thermal annealing should be a topological defect, which keeps the number of carbon atoms a constant. The heptagon-pentagon structure is a typical topological defect. For $N_d > N_s$ and $N_d < N_s$, the introduced defects are atomic vacancies and adatoms respectively. Figure 1(b) shows an enlarged image of several typical point defects, including single-atom vacancies [36] and a flower defect [37]. The flower defect is a grain boundary loop, which is a typical heptagon-pentagon topological defect. Obviously, the method presented here is facile and efficient to introduce a high density of point defects in graphene. This method should also work well for graphene on other metal foils if that the metal foils show temperature dependent carbon solubility. Further experiments using this facile method by depositing graphene monolayer on a metal substrate with pre-define structures maybe helpful to induce periodic point defects into graphene.

In the following, we focus on how the point defects affect the electronic structures of graphene layers. Figure 2(a) shows the reciprocal lattice, the first Brillouin zone, and schematic Dirac cones of graphene monolayer. The charge carriers in graphene possess chirality as ultrarelativistic particles. The chirality suppresses back-scattering of quasiparticles between the two adjacent Dirac cones [5], which is explicitly demonstrated in Fig. 2(b). The fast Fourier transfer (FFT) of a defect-free graphene monolayer shows no signal of intervalley scattering.

Figure 2(c) shows a STM image of graphene monolayer with several flower defects generated by the thermal annealing. The atomic structure around one flower defect was shown in Fig. 2(d). The flower defect, which was predicted to have the lowest energy per dislocation core of any known topological defect in graphene, consists of close-packed heptagon-pentagon rings with a sixfold symmetry [37]. Therefore it is not surprised to introduce many flower defects in our sample by the thermal annealing. A clear peak of the localized state at the flower defect is observed at ~ 270 meV in the STS spectrum, as shown in Fig. 2(e). A $\sqrt{3}\times\sqrt{3}R30°$ interference pattern of carbocyclic rings is observed around the flower defect. This interference pattern in the STM image can also be observed in the FFT of the STM image (the middle set of bright spots in Fig. 2(f)). It is attributed to the elastic scattering process between the two adjacent Dirac cones at K and K′ [20]. It indicates that the flower defect can generate sharp enough scattering potentials to mix the two valleys in graphene.

The center bright region of the FFT image is closely related to the intravalley scattering, which only needs a small moment transfer. One can obtain the local Fermi velocity according to the radius of the center spot of the FFT image [17]. Although the conservation of the sublattice pseudospin suppresses the intravalley scattering in graphene monolayer, the sublattice pseudospin is vulnerable to disorder. The intrinsic curvature, extrinsic ripple (induced by thermal expansion mismatch between graphene and the substrate), lattice defects and deformations of graphene can result in the intravalley scattering. The Fermi velocity of the point-defect-free graphene (Fig. 2(b)) and the graphene with flower defects (Fig. 2(d)) is estimated as $(7.10 \pm 0.50)\times 10^5$ m/s and $(3.47 \pm 0.50)\times 10^5$ m/s respectively [38]. It indicates that the enhanced scattering around the flower defects results in a significant reduction of the Fermi velocity (see Fig. S3 of supplementary material [35] for more experimental results).

Our experiment indicates that the flower defects lead to a localized density of states (DOS), provide sharp enough scattering potentials to mix the two valleys, and result in a reduction of the Fermi velocity. We will show subsequently that these effects are common features of all the point defects in graphene layers. Figure 3 shows several other types of point defects. These point defects show quite different characteristics in their STM images, and, at present, it is very difficult to figure out the atomic structures of all the point defects explicitly. A localized DOS peak is also observed in the STS spectrum of these point defects (see Fig. S4 [35]). The interference pattern, which arises from intervalley scattering, can be observed in both the STM images and the FFT of the STM images of all the point defects, as shown in Fig. 3. The Fermi velocity around these point defects is also reduced significantly to $(3 \sim 6)\times 10^5$ m/s. Our experimental result further demonstrated that deformed graphene structures with size much larger than the lattice spacing of graphene cannot provide sharp enough scattering potentials to mix the two independent valleys. It only leads to the intravalley scattering (see Fig. S5 [35]).

To further explore the effects of point defects on electronic properties of graphene, first-principles calculations using the projector augmented wave pseudopotential method and Perdew-Burke-Ernzerhof exchange-correlation potential [39] implemented in the VASP package [40] have been carried out. The STM simulations are performed using the Tersoff-Hamann model [41]. The supercell of graphene can be divided into

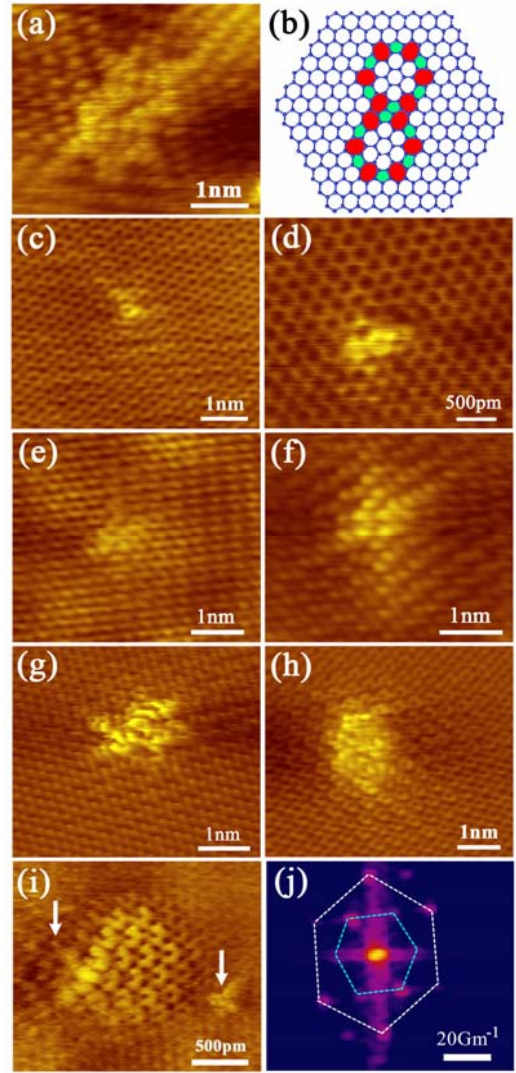

FIG. 3 (color online). (a), (c)-(i) STM images of various point defects in graphene on Rh foil. (b) Schematic structure of the defect in panel (a). (i) Enhanced quantum interference between two defects pointed by the arrows. (j) A typical FFT image of the point defects. The outer six spots correspond to the reciprocal lattice of graphene. The inner six spots represent the intervalley scattering.

two categories, i.e., a 3n×3n (n = 1,2,3…) supercell with valleys K and K′ folded into the Γ point and a non-3n×3n supercell with valleys separated in the momentum space. In the calculation, we select the flower defect, which has a well-defined atomic structure, as a typical structure of the point defects. Two typical configurations, which include one defect in a 8×8 supercell (Fig. 4(a)) and one defect in a 9×9 supercell (Fig. 4(e)), are considered. Fig. 4(b) and Fig. 4(f) are the simulated STM images around the flower defect of the 8×8 and 9×9 supercell respectively. Both of them show a $\sqrt{3}\times\sqrt{3}R30°$ interference pattern of carbocyclic rings around the defect, which consists well with our experimental results (see Figure S6 [35] for more simulated STM images). However, the electronic band structures of the two configurations are quite different, as shown in Fig. 4(c) and 4(g). The 8×8 supercell remains linear band dispersion around the K point, while the 9×9 counterpart opens a gap with the minimum at the Γ point.

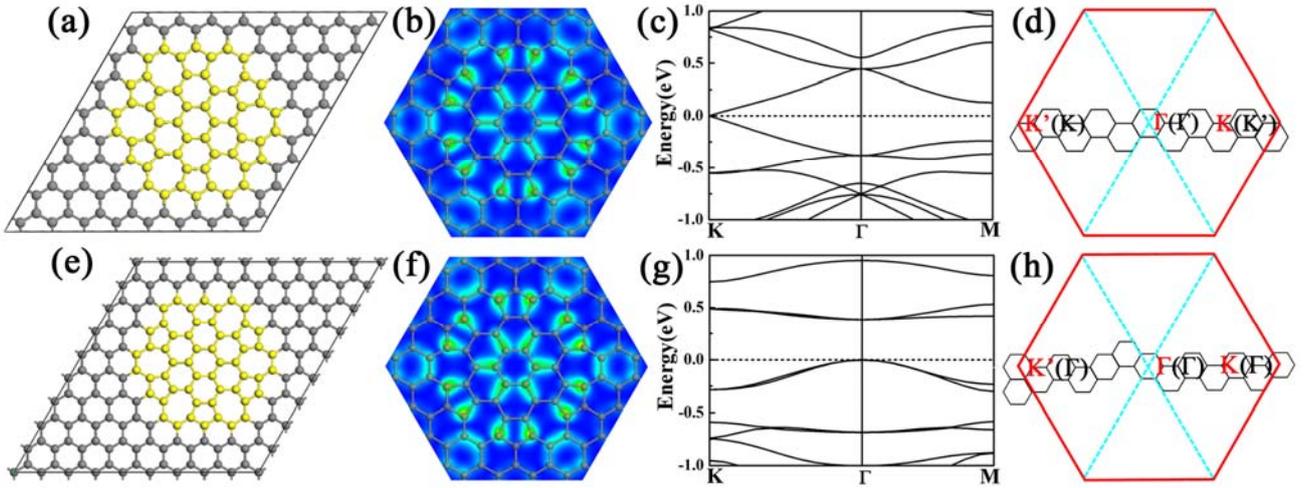

FIG. 4 (color online). (a) and (e) show two typical configurations, which are one flower defect in a 8×8 and one flower defect in a 9×9 supercell, respectively. The atomic structure of the flower defect is shown in yellow. (b) and (f) show the simulated STM images at ~ 500 meV sample bias around the flower defect of panel (a) and (e) respectively. (c) and (g) show electronic band structures of the two configurations in panel (a) and (e) respectively. (d), (h) Brillouin zone for graphene (red borders) and Brillouin zones for the 8×8 supercell and 9×9 supercell (black borders). The center Γ and two inequivalent corners K and K′ of the Brillouin zones are also indicated for the analysis in the text.

The emergence of a significant gap in defect superlattice arises from the fact that the 9×9 supercell hybridizes the two independent Dirac points: both K and K′ of the pristine graphene are folded back to the center Γ of the supercell, as shown in Fig. 4(h). The randomly distributed point defects will weaken the mixture of the two valleys and reduce the gap [42]. As a consequence, the electronic band structure of our sample, which shows complete random distribution of the point defects, should resemble that obtained from a non-3n×3n supercell calculation. For the case of the 8×8 supercell, the K (K′) of the pristine graphene is folded to K′(K) of the 8×8 supercell (Fig. 4(d)). Therefore, the 8×8 supercell has a similar linear band structure as the pristine graphene but with a much reduced Fermi velocity ~ $4.78\times10^5$ m/s, as shown in Fig. 4(c).

Whether the defect superlattice with a 3n×3n supercell will or not open a gap in graphene can be understood more generally from the symmetry point of view [43] and the obtained result should be independent of the type of point defects. It suggests that we can tune the electronic structures of graphene by controlling the density of the point defects and it's possible to realize all-graphene electronics once that the point defects can be patterned into graphene in a controllable way.

In summary, we report a facile method to introduce a high density of point defects in graphene on metal foil and show how the presence of point defects affects the electronic structures of graphene layers. Further experiments should be carried out to control the type of point defects and to pattern the defect superlattice into graphene. We believe that our result may pave a new road towards nanoscale electronic devices based on graphene.

We thank the helpful discussion of X. Q. Lin and J. Ni. This work was supported by the National Natural Science Foundation of China (Grant No. 11004010, No. 10804010, No. 10974019, No. 21073003, No. 51172029, No.11174337, No.11225418, and No. 91121012), the Fundamental Research Funds for the Central Universities, and the Ministry of Science and Technology of China (Grants No. 2011CB921903, No. 2012CB921404, No 2013CB921701, No.2011CBA00100).

§ These authors contributed equally to this paper.
a: ygyao@bit.edu.cn
b,*: helin@bnu.edu.cn.